\documentstyle[12pt,psfig,epsf]{article}
\newcommand{\be}{\begin{equation}}
\newcommand{\ee}{\end{equation}}
\begin{document}
\begin{titlepage}
\title{
{\bf Phase structure of four-dimensional \\
gonihedric spin system }
}%title ends
{\bf
\author{ 
G.Koutsoumbas\\
Physics Department, National Technical University \\
Zografou Campus, 15780 Athens, Greece\\
\vspace{1cm}\\
G.K.Savvidy\\
National Research Center "Demokritos",\\
Ag. Paraskevi, GR-15310 Athens, Greece \\
\vspace{1cm}\\
K.G.Savvidy\\
Princeton University, Department of Physics\\
P.O.BOX 708, Princeton, New Jersey 08544, USA
}%author ends
}
\date{}%in order NOT to write the date
\maketitle
\begin{abstract}
\noindent

We perform Monte Carlo simulations of a gauge invariant spin
system which describes random surfaces with gonihedric action in 
four dimensions. The Hamiltonian is a mixture of one-plaquette 
and additional two-  and three-plaquette interaction 
terms with specially adjusted coupling constants \cite{wegner,sav3}.  
For the system with the large self-intersection coupling constant 
$k$ we observe the second-order phase transition
at temperature $\beta_{c}\simeq 1.75$. The string tension is generated 
by quantum fluctuations as it was  expected theoretically \cite{sav2}.
This result suggests the existence of 
a noncritical string in four dimensions. For smaller values of $k$ the system
undergoes the first order phase transition and for $k$ close to zero  
exhibits a smooth crossover.

\end{abstract}
\thispagestyle{empty}
\end{titlepage}
\pagestyle{empty}

\section{Introduction}
\vspace{.5cm}

The gonihedric string has been defined as a model of random surfaces with 
an action which is proportional to the linear size of the surface 
\cite{ambar,sav1,sav2}

\be
A(M) = \sum_{<ij>} \lambda_{ij}
\cdot \Theta(\alpha_{ij}),~~~~~~~ \Theta(\alpha)= \vert 
\pi - \alpha \vert^{\varsigma} , \label{action}
\ee
where $\lambda_{ij}$ is the length of the edge $<ij>$ of the 
triangulated surface $M_{2}$ and $\alpha_{ij}$ is the dihedral angle 
between two neighbouring triangles of $M$ sharing a common edge $<ij>$.
The angular factor $\Theta$ defines the rigidity of random surfaces 
and for $\varsigma \leq 1$ the angular factor increases sufficiently fast 
near angles $\alpha = \pi$ to suppress transverse fluctuations \cite{sav2}.
The gonihedric action has been defined for self-intersecting surfaces 
as well \cite{sav2}. The  action 
accounts self-intersections of different orders by ascribing 
weights to self-intersections. These weights are proportional 
to the length of the 
intersection multiplied by the angular factor which is equal to the sum of 
all angular factors corresponding to dihedral angles in the 
intersection \cite{sav2}.The coupling constant in front of this term is 
called self-intersection coupling constant \cite{sav3}.
In principle this coupling constant is a free parameter of the theory, 
but if one applies the $continuity~principle$ then one can fix the value
of $k_c =1/2$ \cite{sav2,sav3} (see formulas (4) and (28) in \cite{sav3}).

The model has a number of properties which bring it very close to the Feynman
path integral for a point-like relativistic particle \cite{sav1,sav2}. 
This can be seen from 
(\ref{action}) in the limit when the surface degenerates into a single 
world line, and in that case the action is proportional to the length of the path.
This property of the gonihedric action guarantees that the spike instability, 
which is common to other triangulated random surface theories, 
does not appear here, 
because the action is proportional to the total length of the 
spikes and thus suppresses the corresponding fluctuations 
\cite{ambar,sav1,sav2}.

The other important property of the theory is that 
at the classical level the string tension is equal to zero and 
quarks viewed as open 
ends of the surface are propagating freely without interaction  \cite{sav2}.
This is because the gonihedric action (\ref{action}) 
is equal to the $perimeter$ of the flat Wilson loop  \cite{sav2}.
As it was demonstrated in \cite{sav2}, quantum fluctuations generate a 
nonzero string tension 
\be
\sigma_{quantum}= \frac{d}{a^2}~(1 - ln \frac{d}{\beta}) , \label{tension}
\ee
where $d$ is the dimension of the spacetime, $\beta$ is the coupling constant,
$a$ is a scaling parameter and 

$$\varsigma = \frac{d-2}{d}$$ 
in (\ref{action}).
In the scaling limit $\beta \rightarrow \beta_{c}=d/e$ the string tension 
has a finite limit while the scaling parameter tends to zero as

$$a=(\beta - \beta_{c})^{1/2}, $$
thus the critical exponent $\nu$ is equal to one half \cite{sav2}

\be
\nu = 1/d_{H} = 1/2 , \label{hdim}
\ee
where $d_{H}$ is the Hausdorff dimension.
Thus, although at the tree level the theory describes free quarks with string
tension equal to zero, quantum fluctuations generate nonzero string 
tension and, as a result, the quark confinement \cite{sav2}. The 
gonihedric string may 
consistently describe asymptotic freedom and confinement, as it is expected
to be the case in QCD\footnote{
Different modifications of the action (\ref{action}) have been proposed in 
the literature by adding the area term or the gaussian term. 
In the first case this will cause a nonzero string tension already 
at the classical level and will bring back all problems of the critical 
string when applied to strong interactions. In the second case 
the model becomes simply equivalent to the gaussian model \cite{sav1}
(see formulas (46)-(48)). }.

In addition to the formulation of the theory in the continium space $R^{d}$
the system allows an equivalent representation on Euclidean lattice $Z^{d}$ 
where a surface is associated with a collection of plaquettes.
Lattice spin systems whose interface energy coinsides with the action 
(\ref{action}) have been constructed in arbitrary dimension $d$ \cite{wegner} 
for the self-intersection coupling constant $k=1$ and for an arbitrary 
$k$ in \cite{sav3}. This gives an opportunity for  numerical simulations 
of the corresponding statistical systems in a way which is similar to the 
Monte Carlo simulations of QCD \cite{creutz}.

In three dimensions the corresponding Hamiltonian is equal to \cite{wegner,sav3}

\be
H_{gonihedric}^{3d}=- 2k \sum_{\vec{r},\vec{\alpha}} \sigma_{\vec{r}}
\sigma_{\vec{r}+\vec{\alpha}}
+ \frac{k}{2} \sum_{\vec{r},\vec{\alpha},\vec{\beta}} \sigma_{\vec{r}}
\sigma_{\vec{r}+\vec{\alpha} +\vec{\beta}}
-  \frac{1-k}{2} \sum_{\vec{r},\vec{\alpha},\vec{\beta}} \sigma_{\vec{r}}
\sigma_{\vec{r}+\vec{\alpha}} \sigma_{\vec{r}+\vec{\alpha}+\vec{\beta}}
\sigma_{\vec{r}+\vec{\beta}}, \label{hamil}
\ee
and is a natural extension of the $3D$ Ising model 

$$
H_{Ising}^{3d}= -  \sum_{\vec{r},\vec{\alpha}} \sigma_{\vec{r}}
\sigma_{\vec{r}+\vec{\alpha}}
$$
to the gonihedric case. The  extensions of the $3D$ Ising model have 
been considered in the literature \cite{fan} and 
their  phase structure has been investigated. 
The essential point is that the geometrical 
nature of the gonihedric system specifies the coupling constants and the 
symmetry of the system\footnote{Spoiling of  this fine tuning of coupling 
constants is equivalent to an addition of an area term and thus leads  
to nonzero string tension at the tree level (see previous footnote) 
and "drives" the system to an unwanted class of universality.}.
This is in analogy \cite{wegner,sav3} with supersymmetric systems  
where the symmetry also specifies the coupling constants.

The self-intersection coupling constant $k$ defines the degree of degeneracy 
of the vacuum state \cite{pav,sav4};  
if $k \neq 0$, the degeneracy of the vacuum state is 
equal to $3\cdot 2^N$ for the lattice of the size $N^3$ and is equal to 
$2^{3N}$ when $k=0$. The last case is a sort of supersymmetric point in the 
space of gonihedric Hamiltonians. What is amazing is that not only the 
ground state, but all of the energy levels have the same exponential degeneracy.
This enhanced symmetry at the point $k=0$ 
allows to construct the dual Hamiltonian in three dimensions \cite{sav4,pav}. 
For $k \neq 0$
the degeneracy of energy levels is smaller compared with the vacuum state.
This exponential degeneracy is reminiscent of $spin~ glass~systems$. 

Despite the fact that the system has a rich symmetry, in three dimensions 
the two-point 
correlation function remains as a fundamental observable \cite{bath}

$$
C(\vec{r}) = <~\sigma_{\vec{o}}~\sigma_{\vec{r}}>
$$
and its behaviour at large distances 
is the best indicator of the phase transitions in the 3D gonihedric spin 
system. As usual,  
the most direct indication of the second-order phase transition 
is the growth of the correlation length, which  
in the scaling limit  should tend to infinity. 
The singularities of the energy, of the specific heat and of the generalized 
magnetization are also indicative of the order of the phase transition. 
The generalized magnetization is defined as the vacuum expectation value
of the projection of spin states to the one which corresponds to 
the vacuum spin configuration \cite{bath}

\be
M^{\mu} = <\sum_{\vec r}\sigma^{\mu}_{\vec r}(vac)
\cdot \sigma_{\vec r}> \label{mag}
\ee
where $\sigma^{\mu}_{\vec r}(vac)$ denotes the vacuum spin configurations and 
$\mu =1,2,...,2^{3N}$ for the case $k=0$ and $\mu =1,2,...,3\cdot 2^{N}$ in all
other cases. The exponential degeneracy of the ground state is common to 
gonihedric spin systems and the analogy with spin glasses  
allows to construct the generalized magnetization (\ref{mag}) \cite{bath}.
In the real experiments one can choose some subset of generalized magnetizations
(\ref{mag}) and measure the projection  of virtual states to that subset of 
order parameters \cite{bath}.

The first Monte Carlo simulations \cite{bath} demonstrated that the gonihedric 
system with intersection coupling constant equal to one, $k=1$, undergoes the 
second-order phase transition at $\beta_{c} \approx 0.44$. This happens near 
the critical temperature of the two-dimensional 
Ising model $\beta_{c} = \frac{1}{2} ln(1+\sqrt{2}) \approx 0.44$ \cite{sav4}.
This result follows from the transfer matrix approach \cite{sav4} which 
describes the propagation of the two-dimensional system with the length and 
curvature amplitude. The system has a continuum limit at this temperature and 
describes two-dimensional free Dirac fermion  \cite{sav4}. 
At the same time it was not expected that the critical indices will 
coincide with the ones of the 2D Ising model simply 
because the physical picture of the fermionic string propagation which follows
from the transfer matrix approach \cite{sav4} is different from 
the propagation of the free point like fermion in two dimensions. 
In addition the analytical result (\ref{hdim}) predicts the value $\nu = 0.5$, 
which does not coinside 
with the critical exponent of the 2D Ising model  $\nu = 1$. 
The Monte Carlo simulations \cite{bath} confirm that 
the value of $\nu $ is small, $\nu \approx 0.34$, and is almost twice smaller 
than that for the 3D Ising model where $\nu = 0.63$ \cite{creu}. 
In \cite{des} the value $\nu \approx 0.44$ is obtained. Thus the 
correlation function grows slower near the critical point.
This means that the 3D gonihedric system and the 3D Ising model are in  
different classes of universality.

Our aim is to study the phase structure of the spin systems which 
simulate random 
surfaces with $gonihedric~action$ in four dimensions. As it is well known, 
random surfaces with an $area$ action in four dimensions
can be simulated by the one-plaquette gauge invariant action 
(\ref{area}) \cite{weg}. It should be reminded also 
that this gauge invariant spin system in four dimensions is self-dual 
and the critical temperature is equal to 
$\beta_{c} = \frac{1}{2} ln(1+\sqrt{2})$ \cite{weg}. 
The Monte Carlo simulation of the 
system strongly indicates that the phase transition in 4D $Z_2$ gauge invariant 
spin system is of the first order \cite{creutz1,creutz2}. 

In the next section we shall describe the corresponding gonihedric  
Hamiltonian from \cite{wegner,sav3} and the 
appropriate observables. The results of the Monte Carlo simulations are 
presented in the fourth section. For the system with the self-intersection 
coupling constant $k > k^{'}_c = 1/2$ we observe the 
second-order phase transition
at temperature $\beta_{c}\simeq 1.75$. The string tension is generated 
by quantum fluctuations as it was  expected theoretically \cite{sav2}.
This result suggests the existance of 
a noncritical string theory in four dimensions.  

For smaller values of $k$ the system
undergoes a first-order phase transition at temprature $\beta_{c} \geq 2$. 
The passage from second-order phase transition to the first-order one is 
caused by the {\it condensation of self-intersections}. 
Finally, for $k < k^{''}_c = 1/6$ the system exhibits a smooth crossover. 
Thus two critical 
values of $k$ separate three regions of different critical behaviour and 
demonstrate the essential dependence of the phase structure on the 
intersection coupling constant $k$.

\section{Gonihedric spin system in four dimensions}

The system of random surfaces with gonihedric action can be formulated 
not only in three dimensions, but also in any dimension \cite{wegner}. 
Similar to the
three-dimensional case , the surface can be associated with a collection of 
plaquettes 
on a lattice and the interaction between spins can be organized in a 
way that the surfaces of interface will have the 
gonihedric action \cite{wegner}.

In four dimensions the gonihedric Hamiltonian essentially differs from the 
three-dimensional one (\ref{hamil}) because now it represents a 
gauge invariant spin 
system. The gauge invariant Hamiltonian in four dimensions has 
the form \cite{wegner}

\be
H^{4d}_{gonihedric}= -\frac{4}{g^{2}} \sum_{\{plaquettes\}}
(\sigma\sigma\sigma\sigma) + \frac{1}{4g^{2}} 
\sum_{\{right~angle~plaquettes\}}
(\sigma\sigma\sigma\sigma_{\alpha})^{rt}
(\sigma_{\alpha}\sigma\sigma\sigma), \label{goni}
\ee
where $g^{2}$ is the gauge coupling constant and the independent spin variables 
$\sigma_{ij}$ should be attached to the centers of the edges $<i,j>$ of the 
four-dimensional lattice. 

The 
Hamiltonian which simulates random surfaces with $area$ action in 
four dimensions
is well known  \cite{weg} and represents a gauge invariant spin system with 
one-plaquette interaction term

\be
H^{4d}_{area}= -\frac{1}{g^{2}} \sum_{\{plaquettes\}}
(\sigma\sigma\sigma\sigma) .\label{area}
\ee
Thus the gonihedric Hamiltonian (\ref{goni}) in addition to the "ferromagnetic"
interaction (\ref{area}) on each elementary plaquette has the 
"antiferromagnetic" 
interaction between two plaquettes which form a right angle \cite{wegner}

$$J_{plaquettes}= - 16\cdot J_{rt~plaquettes} =
\frac{4}{g^{2}}.$$
The partition function corresponding to (\ref{goni}) is equal to 

$$Z(\beta) = \sum_{\{\sigma_{\vec r}\}} e^{-\beta~g^{2} H/4}. $$
These formulas completely define the gonihedric system in four dimensions 
and allow to simulate random surfaces on a four-dimensional lattice. It  
is important to stress that both systems (\ref{goni}) and (\ref{area}) 
are of a geometrical nature because in the first case the 
amplitudes are proportional
to the $linear~size$ of a surface and in the second case they are 
proportional to the $area$. The other difference between these 
two systems is that
in the gonihedric case (\ref{goni}) the self-intersections of the 
surface can be propertly counted \cite{sav2,sav3}. 
There are various formulations of self-avoiding random 
surfaces on a lattice \cite{maritan}. The main difference between these 
models and the gonihedric 
case is that the energy ascribed to self-intersections  essentially depends
on the  geometry of intersecting plaquettes.

Indeed, the  essential property of the spin realization of 
the gonihedric system (\ref{goni}) is that it properly counts the 
self-intersections \cite{sav2}. On a four-dimensional lattice 
a two-dimensional 
closed surface can have self-intersections of different orders because at a 
given edge one can have intersections of four or six plaquettes 
\cite{wegner}. The energy ascribed to a self-intersection essentially
depends on a configuration of plaquettes in the intersection \cite{sav2}, 
and there are two topologically different 
configurations of plaquettes with four intersecting plaquettes and only 
one with six intersecting plaquettes. 
The corresponding energies as they are defined
by the Hamiltonian (\ref{goni}) ($g^{2}=1$) are equal to \cite{wegner}

\be
\epsilon_{2} =1,~~\bar{\epsilon}_{4}=4,~~
\bar{\bar{\epsilon}}_{4}=5,~~\epsilon_{6}=12 \label{weight}
\ee
and are equal to the number of plaquettes which intersect at a 
right angle.
The total energy of the surface which has $n_{2}$ edges with two intersecting
plaquettes at a right angle - "simple edges"\footnote{edges with  $two$ 
intersecting plaquettes at a right angle}, $\bar{n}_{4}$ and 
$\bar{\bar{n}}_{4}$ edges with
intersection of four plaquettes and $n_{6}$ with six plaquettes is equal to

\be
\epsilon= n_{2} + 4 \bar{n}_{4} + 5 \bar{\bar{n}}_{4} + 12 n_{6} \label{energy}
\ee
In the article \cite{sav3} the intersection coupling constant $k$ was 
introduced to control the intensity of self-intersections. In that case 
the Hamiltonian  has the form \cite{sav3}

$$
H^{\kappa}_{gonihedric}= -\frac{5\kappa-1}{g^{2}} \sum_{\{plaquettes\}}
(\sigma\sigma\sigma\sigma) + \frac{\kappa}{4g^{2}} 
\sum_{\{right~angle~plaquettes\}}
(\sigma\sigma\sigma\sigma_{\alpha})^{rt}
(\sigma_{\alpha}\sigma\sigma\sigma)$$
\be
-\frac{1-\kappa}{8g^{2}} 
\sum_{\{triples~of~right~angle~plaquettes\}}
(\sigma\sigma\sigma\sigma_{\alpha})^{rt}
(\sigma_{\alpha}\sigma\sigma\sigma_{\beta})^{rt}
(\sigma_{\beta}\sigma\sigma\sigma)  .               \label{gonih}
\ee
As it can be seen from the last expression in contrast to  
the case $\kappa=1$ (\ref{goni}) there appears   
a new three-plaquette interaction 
term\footnote{ The coefficient in front of this term is twice smaller 
than that in \cite{sav3} because for convinience we take all 
symmetric combinations of the plaquette triples inside the 3d cube.}.

The corresponding energies ascribed to  
self-intersections are equal to \cite{sav3}

$$\epsilon_{2} =1,~~\bar{\epsilon}_{4} = 
4k,~~\bar{\bar{\epsilon}}_{4}=6k-1,~~\epsilon_{6}=12k$$
and coinside with the weights (\ref{weight}) when $k=1$. 
The total energy of the surface in this general case is   

\be
\epsilon= n_{2} + 4k \bar{n}_{4} + (6k-1) \bar{\bar{n}}_{4} + 12k n_{6}
\label{energyk}
\ee
and  reduces to  (\ref{energy}) when $k=1$. 

Unfortunately the method used for the construction of the 
transfer matrix in three dimensions \cite{sav2} can not be used in 
four dimensions and 
we face an even a more complicated problem.

\section{Observables}

In  pure gauge theories the magnetization is always equal to zero, 
$<\sigma> =0$, therefore this quantity can not be the order 
parameter. In other words the system does not have local order parameters 
to distinguish the phases and only gauge invariant 
products of spin variables over closed loops can be nonzero 

\be
W(C) ~~=~~ <~\prod_{C} \sigma~> \label{wegnerloop}
\ee
and, as it was pointed out by Wegner \cite{weg}, can separate  the phases 
of the system.  The perimeter and the 
area law of (\ref{wegnerloop}) can distinguish the ordered and disordered 
phases  \cite{weg} (see formulas (3.34) - (3.39) in \cite{weg}). 
The same conclusion is correct for the gonihedric 
systems (\ref{goni}) and  (\ref{gonih}), therefore only different loop 
products are to be considered. 

To define the phase structure of the four-dimensional gonihedric gauge 
system one can measure different observables. The simplest one is the 
total energy per edge which can be expressed as a combination of different 
loop products of spins inside a three-dimensional cube 
\cite{sav3}

\be
H^{tot}_{edge}=~~ 3k~~-~~\frac{5\kappa-1}{4} \sum_{\{six\}}
(\sigma\sigma\sigma\sigma)~~ +~~ \frac{\kappa}{4} 
\sum_{\{twelve\}}
(\sigma\sigma\sigma\sigma_{\alpha})^{rt}
(\sigma_{\alpha}\sigma\sigma\sigma)   \label{peredge}
\ee
$$-~\frac{1-\kappa}{8} 
\sum_{\{twelve\}}
(\sigma\sigma\sigma\sigma_{\alpha})^{rt}
(\sigma_{\alpha}\sigma\sigma\sigma_{\beta})^{rt}
(\sigma_{\beta}\sigma\sigma\sigma) .$$
Here the first term denotes the product of spins around one 
plaquette in a 
3d cube, the second one - a product of spins around two 
perpendicular plaquettes and the 
last one  - a product of spins around three perpendicular plaquettes
inside a 3d cube of the four-dimensional lattice.

\begin{figure}
\centerline{\hbox{\psfig{figure=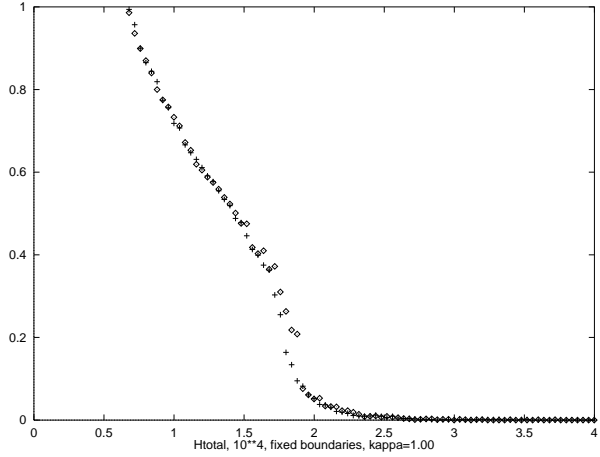,height=6cm,angle=-90}}}
\caption[fig1]{The average energy 
per edge (\ref{peredge}) as a function of $\beta$ . The system was heated 
and then cooled on steps of 0.001, points are plotted every 
fortieth step, $k=1$.}
\label{fig1}
\end{figure}

The advantage of this formulation of the theory is that one can 
measure the part of the surface 
energy which corresponds only to a $simple~edge$ which is proportional 
to $n_{2}$

\be
16~ H^{simple}_{edge}=~~ 6~~ + ~~\sum_{\{six\}}
(\sigma\sigma\sigma\sigma)~~-~~2 
\sum_{\{three\}}(\sigma\sigma\sigma\sigma)^{\vert \vert}
(\sigma\sigma\sigma\sigma)          \label{single}
\ee
$$-~~\sum_{\{twelve\}}
(\sigma\sigma\sigma\sigma_{\alpha})^{rt}
(\sigma_{\alpha}\sigma\sigma\sigma_{\beta})^{rt}
(\sigma_{\beta}\sigma\sigma\sigma).$$
The separation of the simple edge energy allows to compute the part of 
the surface energy which corresponds to self-intersections 

$$H^{self-intersections}_{edge}~~=~~  H^{tot}_
{edge}~~ -~~ H^{simple}_{edge}$$
One can express this observables directly in terms of numbers of intersections

$$
H^{tot}_{edge} \approx 
n_{2} + 4k \bar{n}_{4} + (6k-1) \bar{\bar{n}}_{4} + 12k n_{6}
$$
$$
H^{simple}_{edge} \approx n_{2}
$$
$$H^{self-intersections}_{edge} \approx 
4k \bar{n}_{4} + (6k-1) \bar{\bar{n}}_{4} + 12k n_{6}
$$
The other important observable is the surface area which can be computed 
by using one-plaquette product

\be
\Sigma_{edge} = 3 - \frac{1}{2} \sum_{\{six\}} 
(\sigma\sigma\sigma\sigma), \label{onep}
\ee
where $(\sigma\sigma\sigma\sigma) = P$.
The advantage to work with these observables is that they allow to 
control the inportance of the self-intersections in our attempt to find
a nontrivial theory in four dimensions.

\begin{figure}
\centerline{\hbox{\psfig{figure=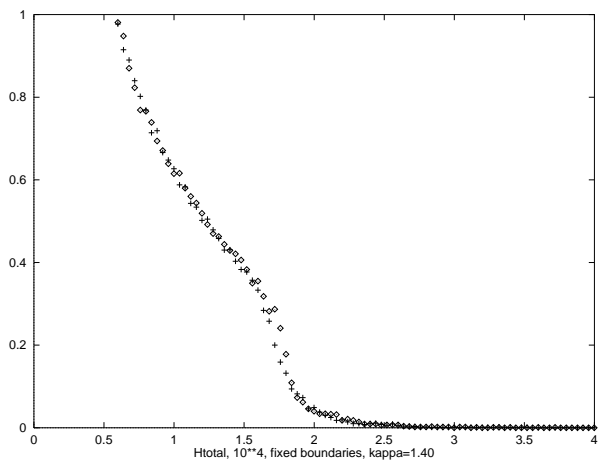,height=6cm,angle=-90}}}
\caption[fig2]{Total energy on edge (\ref{peredge}), $k=1.4$}
\label{fig2}
\end{figure}
\begin{figure}
\centerline{\hbox{\psfig{figure=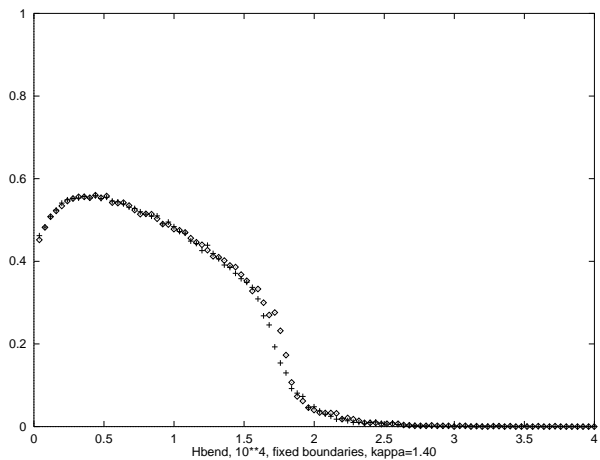,height=6cm,angle=-90}}}
\caption[fig3]{Simple edges (\ref{single}), $k=1.4$}
\label{fig3}
\end{figure}
\begin{figure}
\centerline{\hbox{\psfig{figure=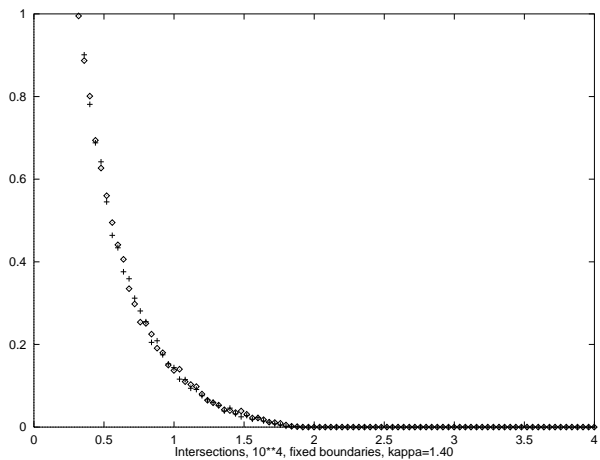,height=6cm,angle=-90}}}
\caption[fig4]{Self-intersection edges, $k=1.4$}
\label{fig4}
\end{figure}
\begin{figure}
\centerline{\hbox{\psfig{figure=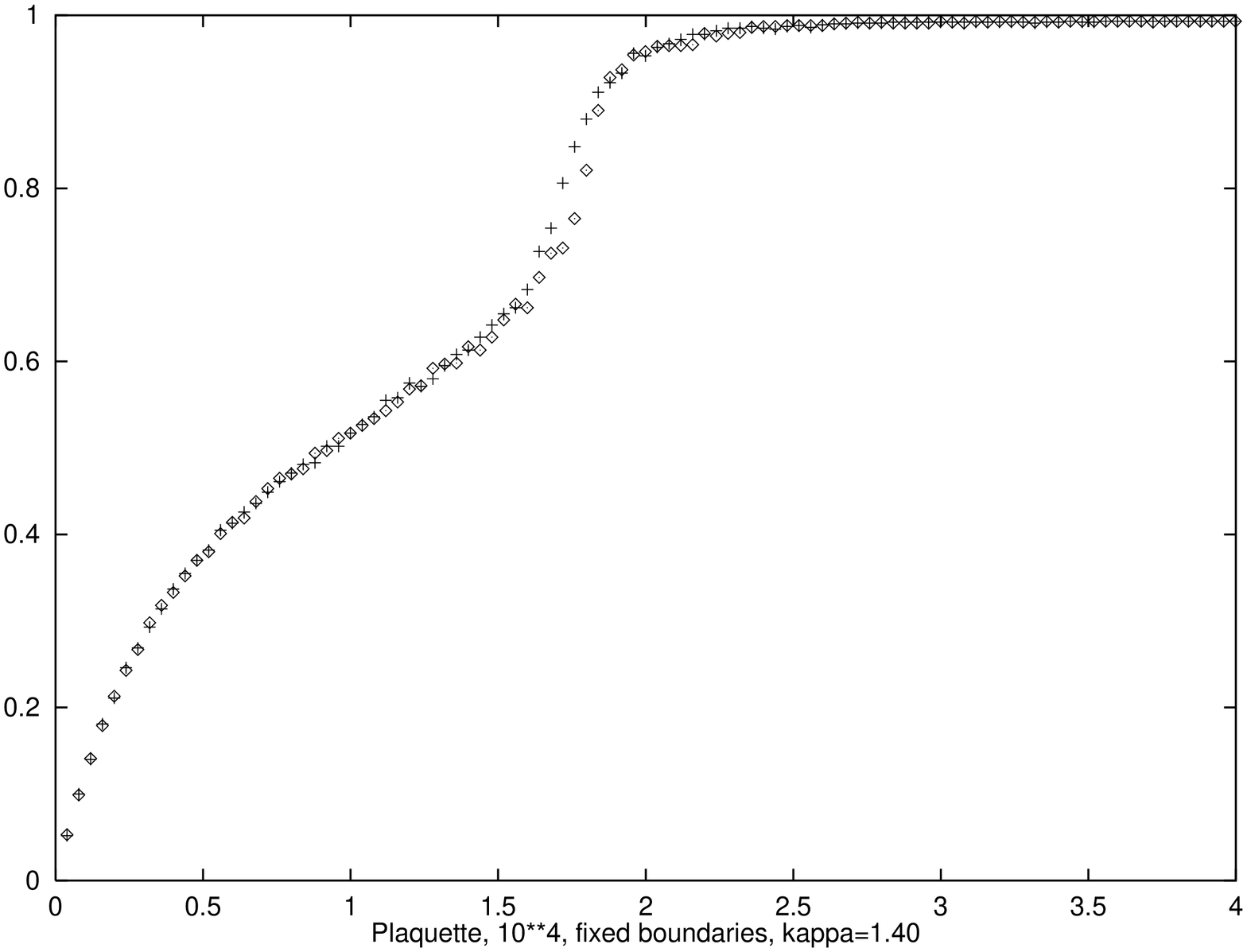,height=6cm,angle=-90}}}
\caption[fig5]{Plaquette, $k=1.4$}
\label{fig5}
\end{figure}

\section{Monte Carlo simulations}

As it is well known, random surfaces with an area action can be simulated by the 
gauge invariant action (\ref{area}) \cite{weg}. It should be reminded 
that this gauge invariant spin system in four dimensions is self-dual 
and the critical temperature is equal to 
$\beta_{c} = \frac{1}{2} ln(1+\sqrt{2})$ \cite{weg}. 
The Monte Carlo simulation of the 
system strongly indicates that the phase transition in 4D $Z_2$ gauge invariant 
spin system is of the first order \cite{creutz1,creutz2}. This has been 
indicated by measuring one-plaquette average (\ref{onep}) in the thermal cycle.
A thermal cycle
of the statistical system  provides a general overview of its phase structure 
and is very helpful in defining the regions of the phase transitions 
\cite{creutz1}.
Clear hysteresis in 4D $Z_2$ gauge invariant spin system
is strongly indicative of the first-order phase transition \cite{creutz1}. 
The hysteresis 
is due to the metastability of the ordered phase at high temperature and of the
disordered phase at low temperature.

The similar Monte Carlo simulations of the gonihedric system 
(\ref{goni}) and  (\ref{gonih}) on the lattices 
of the sizes $4^4$,~$6^4$,~$8^4$,~and~ $10^{4}$ demonstrate that the 
system has three regions of different critical behaviour which are 
characterized 
by the value of $k$. The thermal cycles with steps $\delta \beta = 0.001$ 
have been performed in a large interval of the 
self-intersection coupling constant $k$ from zero to four 
and we see that the phase structure essentially depends on $k$. 

For the system with the self-intersection coupling constant $k > k^{'}_c = 1/2$ 
we observe the second-order phase transition
at temperature $\beta_{c}\simeq 1.75$ ~~ (see Figures 1, 2).
figures 1, 2 show the energy per edge as a function
of $\beta$ in Monte Carlo simulations. 
The energy is a continuous function  of temperature for large values of  
the self-intersection coupling constant $k$. 
The critical temperature does not actually 
depend on $k$. The small hysteresis loop still seen on the figures 
is a remnant of the loop which shrinks 
when we change $\delta \beta$ from $0.1$ to $0.001$ and pass to large 
lattices, this is indicative for the second-order phase transition. 
The behaviour of the simple edges, self-intersections and plaquette 
averages are shown respectively on figures 3,4 and 5. From the figures 3 and 4 
one can conclude that $only$ at high temperature $\beta \leq 0.5 $ the 
energy is concentrated on the self-intersections and that at the 
phase transition point the {\it amount of energy on the self-intersections is 
negligible}. The behaviour of the plaquette average on figure.5 
shows that the area of the random surfaces scales at the critical 
point and signals that the string tension is generated by quantum fluctuations, 
as it is expected theoretically (see (\ref{tension})). 
The Creutz ratio \cite{creutz} will be presented in a separate place.

\begin{figure}
\centerline{\hbox{\psfig{figure=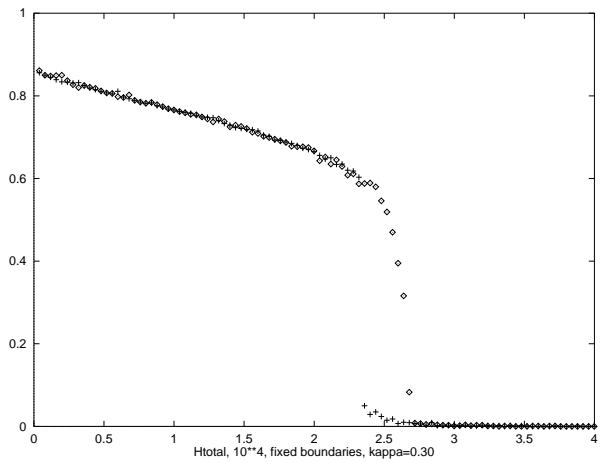,height=6cm,angle=-90}}}
\caption[fig6]{Hysteresis loop for total energy on edge, $k=0.3$}
\label{fig6}
\end{figure}
\begin{figure}
\centerline{\hbox{\psfig{figure=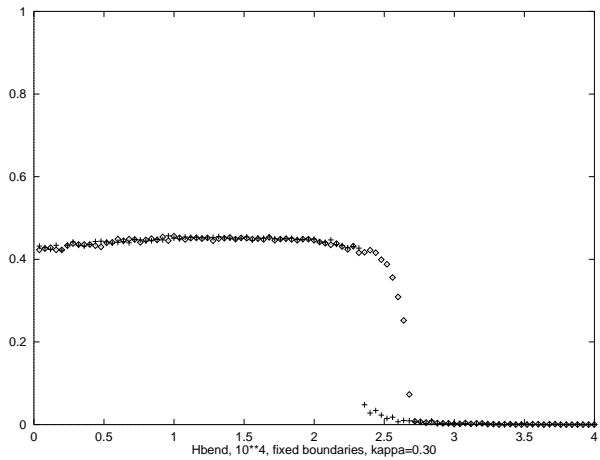,height=6cm,angle=-90}}}
\caption[fig7]{Hysteresis loop for simple edges, $k=0.3$}
\label{fig7}
\end{figure}
\begin{figure}
\centerline{\hbox{\psfig{figure=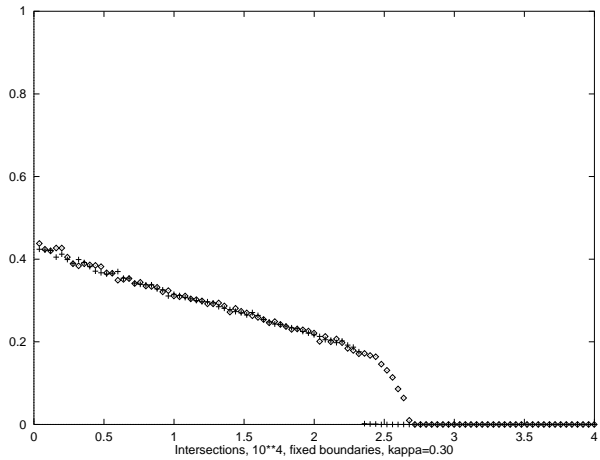,height=6cm,angle=-90}}}
\caption[fig8]{Hysteresis loop for self-intersection edges, $k=0.3$}
\label{fig8}
\end{figure}
\begin{figure}
\centerline{\hbox{\psfig{figure=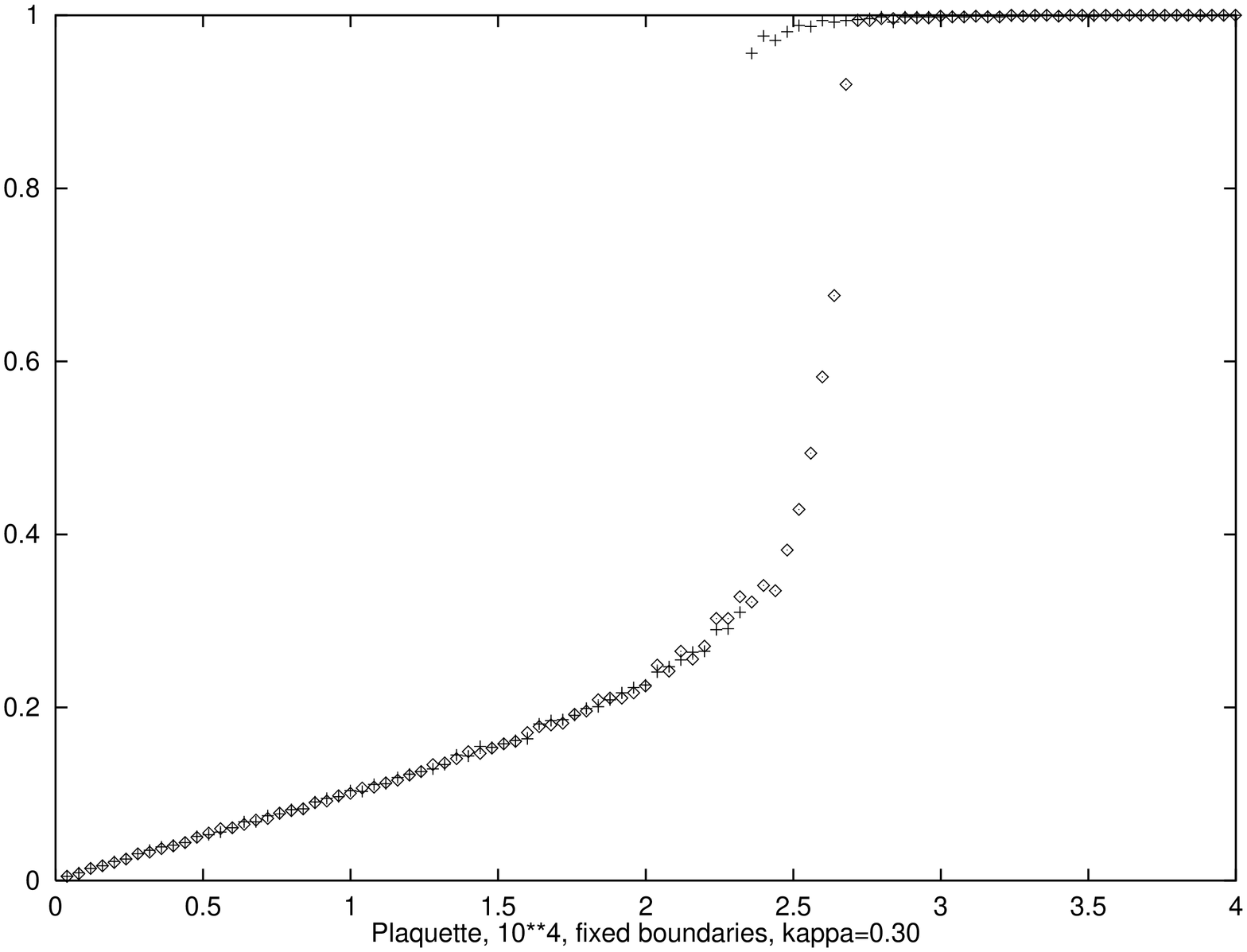,height=6cm,angle=-90}}}
\caption[fig9]{Hysteresis loop for plaquette, $k=0.3$}
\label{fig9}
\end{figure}
For smaller values of $k$ the system
undergoes a first order phase transition at temeprature $\beta_{c} \geq 2$ 
(see figures 6, 7 and 8 ). The energy is discontinuous function of temperature,  
see figure 6. The critical temperature depends on $k$ and is equal to 
$\beta_{c}\simeq 2.5$ for $k=0.3$. At the phase transition point the amount 
of energy on self-intersections considerably increases ( compare figures 2, 3, 4 
and figures 6, 7, 8) and is comparable with the energy on simple edges ( see 
figures 7 and 8). Thus the passage from the second-order phase transition to the 
first-order one  is caused by the {\it condensation of self-intersections}. 
The reason why $k^{'}_c =1/2$ 
is a critical value for the self-intersection coupling constant 
is that at this value the self-intersection energy can be equivalently 
considered 
as being a sum of simple edges. The self-intersection dissolves 
into a number of  simple edges and there is 
no suppression any more of self-intersections compared with the simple edges. 
The plaquette average shows strong discontinuity, see Fig.9, and the area does 
not scale in this case.

Finally, for $k < k^{''}_c = 1/6$ the system exhibits a smooth crossover,  
see Fig.10, and surfaces "evaporate" at all temperatures.

\begin{figure}
\centerline{\hbox{\psfig{figure=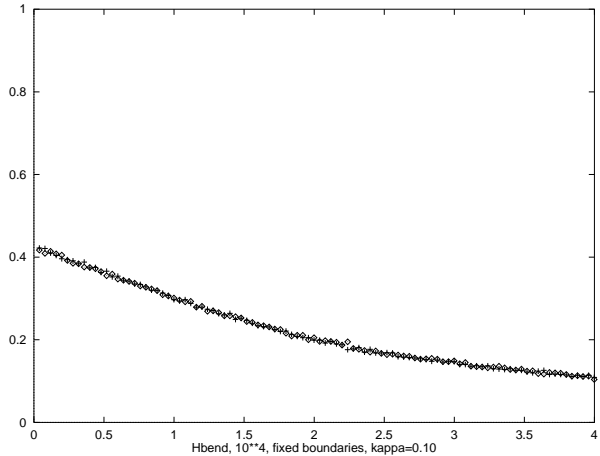,height=6cm,angle=-90}}}
\caption[fig10]{Simple edges, $k=0.1$}
\label{fig10}
\end{figure}

\section{Conclusion}

In conclusion we would like to stress that for the system with the large
self-intersection 
coupling constant $k$ we observe the 
second-order phase transition
at temperature $\beta_{c}\simeq 1.75$. The string tension is generated 
by quantum fluctuations as it was  expected theoretically \cite{sav2}.
This result suggests the existence of 
a noncritical string theory in four dimensions.

\vfill

\begin{thebibliography}{99}

\bibitem{wegner}G.K.Savvidy and F.J.Wegner. Nucl.Phys.B413(1994)605.

\bibitem{sav3}G.K. Savvidy and K.G. Savvidy. Phys.Lett. B324 (1994) 72

\bibitem{weg}F.J. Wegner, J. Math. Phys.12 (1971) 2259

\bibitem{creutz1}M.Creutz,  L. Jacobs and C. Rebbi
Phys.Rev.Lett. 42 (1979) 1390

\bibitem{fan} C.Fan and F.Y.Wu. Phys.Rev.179 (1969)560 \\
F.Y.Wu. Phys.Rev.183 (1969) 604 \\
W. Selke, Physics Reports 170 (1988) 213 \\
M.E. Fisher and W.Selke, Phys. Rev. Lett. 44(1980) 1502 \\
E.I. Dinaburg and Ya.G. Sinai, Comm.Math.Phys. 98 (1985) 119 \\
D.P. Landau, K. Binder, Phys. Rev. B 31 (1985) 5946 \\
A.Cappi, P.Colangelo, G.Gonella and A.Maritan Nucl.Phys. B370 (1992) 659


\bibitem{maritan}A.Maritan and C.Omero. Phys.Lett. B109 (1982) 51\\
T.Sterling and J.Greensite. Phys.Lett. B121 (1983) 345 \\
B.Durhuus,J.Fr\"ohlich and T.Jonsson. Nucl.Phys.B225 (1983) 183 \\
T.Hofs\"ass and H.Kleinert. Phys.Lett. A102 (1984) 420 \\
M.Karowski and H.J.Thun. Phys.Rev.Lett. 54 (1985) 2556\\
F.David. Europhys.Lett. 9 (1989) 575

\bibitem{ambar}R.V. Ambartzumian, G.K. Savvidy , K.G. Savvidy
and G.S. Sukiasian. Phys. Lett. B275 (1992) 99

\bibitem{sav1} G.K. Savvidy and K.G. Savvidy. Int. J. Mod. Phys. 
A8 (1993) 3993.

\bibitem{sav2}G.K. Savvidy and K.G. Savvidy. Mod.Phys.Lett. A8 (1993) 2963.

\bibitem{durhuus}B.Durhuus and T.Jonsson. Phys.Lett. B297 (1992) 271

\bibitem{baillie}C.F.Baillie and D.A.Johnston. 
Phys.Rev D 45 (1992) 3326


\bibitem{bal}R.Balian, J.M.Drouffe and C.Itzykson. Phys.Rev. D10  (1974) 3376;\\ 
Phys.Rev. D11  (1975) 2098, 2104


\bibitem{creutz} L.Kadanoff et al., Rev.Mod.Phys. 39 (1967) 395\\
K.G.Wilson and J.Kogut, Phys.Rev. 12C (1974) 75 \\
M.Creutz, Quarks, gluons and lattices. (Cambridge University
Press,Cambridge 1983)\\
A.Polyakov. Gauge fields and String. (Harwood Academic Publishers, 1987)\\
J.Zinn-Justin. Quantum field theory and critical phenomena. (Oxford University
Press, New York 1989)

\bibitem{creutz2}G. Bhanot and M.Creutz Phys.Rev. D21 (1980) 2892

\bibitem{sav4}G.K. Savvidy and K.G. Savvidy. Phys.Lett. B337 (1994) 333\\
Mod.Phys.Lett. A11 (1996) 1379.

\bibitem{pav}G.K. Savvidy, K.G. Savvidy and P.G. Savvidy. 
Phys.Lett. A221 (1996) 233

\bibitem{bath} G.K.Bathas, E.Floratos, G.K.Savvidy and K.G.Savvidy.
Mod.Phys.Lett. A10 (1995) 2695

\bibitem{wegpie}R. Pietig and F.J. Wegner. Nucl.Phys. B466 (1996) 513

\bibitem{des}M.Baig, D.Espriu, D.Johnson and R.K.P.C.Malmini, 
J.Phys. A30 (1997) 407; \\ hep-lat/9703008

\bibitem{pellizola}E.Cirillo, G.Gonnella and A.Pellizzola, cond-mat/9612001

\bibitem{creu}G.Bhanot, M.Creutz and H.Neuberger
Nucl.Phys.B235 (1984) 417;\\
M.Creutz. Ann. Phys.167 (1986) 62

\bibitem{savvidy1}G.K.Savvidy and N.G.Ter-Arutyunian, On the Monte-Carlo
Simulation of Physical Systems, J.Comput.Phys. 97, 566 (1991);\\
Preprint EPI-865(16)-86, Yerevan Jun.1986.

\bibitem{akopov1}N.Z.Akopov,G.K.Savvidy and N.G.Ter-Arutyunian, Matrix 
Generator of Pseudorandom Numbers, J.Comput.Phys.97, 573 (1991);\\
Preprint EPI-867(18)-86, Yerevan Jun.1986; 

\end{thebibliography}
\end{document}